\newcommand{\CLASSINPUTtoptextmargin}{0.75in}%
\newcommand{\CLASSINPUTbottomtextmargin}{1in}%
\newcommand{\CLASSINPUTinnersidemargin}{0.63in}%
\newcommand{\CLASSINPUToutersidemargin}{0.63in}%
\documentclass[conference,10pt,letterpaper]{IEEEtran}%
\usepackage{multirow}
\usepackage[none]{hyphenat}
\usepackage{float}
\usepackage{subfig}
\usepackage{dblfloatfix}

\usepackage{t1enc}
\usepackage{times}
\usepackage{comment}
\usepackage{amsmath,amsfonts}
\usepackage{amssymb}
\usepackage[utf8]{inputenc}
\usepackage{graphicx}
\usepackage{cite}
\usepackage[table]{xcolor}
\usepackage{mathrsfs}
\usepackage{nccmath}
\usepackage{mathtools}
\usepackage{bm}
\usepackage{lipsum}
\usepackage{hyperref}
\usepackage{xfrac}
\usepackage{nicefrac}
\usepackage{array}
\usepackage{makecell,tabularx}
\usepackage{threeparttable}
\usepackage{algorithmic}
\usepackage{algorithm}
\usepackage{textcomp}
\usepackage{url}
\usepackage{verbatim}

\newcolumntype{g}{>{\columncolor{gray!50}}c} 
\newcolumntype{P}[1]{>{\centering\arraybackslash}p{#1}} 
\newcolumntype{M}[1]{>{\centering\arraybackslash}m{#1}}
\newcolumntype{N}{@{}m{0pt}@{}}

\makeatletter

\def\@RFICauthorblockNAMEstyle{\normalfont\RFICauthorsize}
\def\@RFICauthorblockAFFILstyle{\normalfont\RFICaffilsize}
\def\@RFICauthorblockEMAILstyle{\normalfont\RFICaffilsize}
\def\RFICauthorblockNAME#1{%
\relax\@RFICauthorblockNAMEstyle%
#1%
}%
\def\RFICauthorblockAFFIL#1{%
\relax\@RFICauthorblockAFFILstyle%
\vskip\@IEEEauthorblockAtopspace
#1%
}%
\def\RFICauthorblockEMAIL#1{%
\relax\@RFICauthorblockEMAILstyle%
\vskip\@IEEEauthorblockAtopspace
#1%
}%
\newcommand{\RFICauthor}[1]{%
\ifIsBlindReviewVersion%
\author{\phantom{\parbox{\textwidth}{\center\relax#1}}}%
\else%
\author{\parbox{\textwidth}{\center\relax#1}}%
\fi%
}%
\newif\ifIsBlindReviewVersion
\def\RFICthispaperforblindreview{\IsBlindReviewVersiontrue}
\def\RFICthispaperforfinalpublication{\IsBlindReviewVersionfalse}
\def\@maketitle{\newpage
\bgroup\par\addvspace{0.5\baselineskip}\centering%
\ifCLASSOPTIONtechnote
   {\bfseries\large\@IEEEcompsoconly{\sffamily}\@title\par}\vskip 1.3em{\lineskip .5em\@IEEEcompsoconly{\sffamily}\@author
   \@IEEEspecialpapernotice\par{\@IEEEcompsoconly{\vskip 1.5em\relax
   \@IEEEtitleabstractindextextbox{\@IEEEtitleabstractindextext}\par
   \hfill\@IEEEcompsocdiamondline\hfill\hbox{}\par}}}\relax
\else
   \vskip0.2em{\RFICtitlesize\ifCLASSOPTIONtransmag\bfseries\LARGE\fi\@IEEEcompsoconly{\sffamily}\@IEEEcompsocconfonly{\normalfont\normalsize\vskip 2\@IEEEnormalsizeunitybaselineskip
   \bfseries\Large}\@title\par}\vskip1.0em\par
   \ifCLASSOPTIONconference%
      {\@IEEEspecialpapernotice\mbox{}\vskip\@IEEEauthorblockconfadjspace%
       \mbox{}\hfill\begin{@IEEEauthorhalign}\@author\end{@IEEEauthorhalign}\hfill\mbox{}\par}\relax
   \else
      \ifCLASSOPTIONpeerreviewca
         {\@IEEEcompsoconly{\sffamily}\@IEEEspecialpapernotice\mbox{}\vskip\@IEEEauthorblockconfadjspace%
          \mbox{}\hfill\begin{@IEEEauthorhalign}\@author\end{@IEEEauthorhalign}\hfill\mbox{}\par
          {\@IEEEcompsoconly{\vskip 1.5em\relax
           \@IEEEtitleabstractindextextbox{\@IEEEtitleabstractindextext}\par\hfill
           \@IEEEcompsocdiamondline\hfill\hbox{}\par}}}\relax
      \else
         \ifCLASSOPTIONtransmag
           {\@IEEEspecialpapernotice\mbox{}\vskip\@IEEEauthorblockconfadjspace%
            \mbox{}\hfill\begin{@IEEEauthorhalign}\@author\end{@IEEEauthorhalign}\hfill\mbox{}\par
           {\vspace{0.5\baselineskip}\relax\@IEEEtitleabstractindextextbox{\@IEEEtitleabstractindextext}\vspace{-1\baselineskip}\par}}\relax
         \else
           {\lineskip.5em\@IEEEcompsoconly{\sffamily}\sublargesize\@author\@IEEEspecialpapernotice\par
           {\@IEEEcompsoconly{\vskip 1.5em\relax
            \@IEEEtitleabstractindextextbox{\@IEEEtitleabstractindextext}\par\hfill
            \@IEEEcompsocdiamondline\hfill\hbox{}\par}}}\relax
         \fi
      \fi
   \fi
\fi\par\addvspace{0.0\baselineskip}\egroup}

\def\RFICtitlesize{\@setfontsize{\RFICtitlesize}{18}{21pt}}
\def\RFICauthorsize{\@setfontsize{\RFICauthorsize}{12}{13pt}}
\def\RFICaffilsize{\@setfontsize{\RFICaffilsize}{12}{13pt}}
\def\RFICcaptionsize{\@setfontsize{\RFICcaptionsize}{8}{9pt}}
\def\RFICbibsize{\@setfontsize{\RFICbibsize}{8}{9pt}}

\def\@IEEEauthorblockNstyle{\RFICauthorsize\@IEEEcompsocnotconfonly{\sffamily}\@IEEEcompsocconfonly{\large}}
\def\@IEEEauthorblockAstyle{\RFICaffilsize\@IEEEcompsocnotconfonly{\sffamily}\@IEEEcompsocconfonly{\itshape}\@IEEEcompsocconfonly{\large}}
\def\@IEEEauthordefaulttextstyle{\RFICauthorsize\@IEEEcompsocnotconfonly{\sffamily}\sublargesize}

\def\thebibliography#1{\section*{\refname}%
    \addcontentsline{toc}{section}{\refname}%
    \RFICbibsize\@IEEEcompsocconfonly{\small}\vskip 0.3\baselineskip plus 0.1\baselineskip minus 0.1\baselineskip
    \list{\@biblabel{\@arabic\c@enumiv}}%
    {\settowidth\labelwidth{\@biblabel{#1}}%
    \leftmargin\labelwidth
    \advance\leftmargin\labelsep\relax
    \itemsep \IEEEbibitemsep\relax
    \usecounter{enumiv}%
    \let\p@enumiv\@empty
    \renewcommand\theenumiv{\@arabic\c@enumiv}}%
    \let\@IEEElatexbibitem\bibitem%
    \def\bibitem{\@IEEEbibitemprefix\@IEEElatexbibitem}%
\def\newblock{\hskip .11em plus .33em minus .07em}%
\ifCLASSOPTIONtechnote\sloppy\clubpenalty4000\widowpenalty4000\interlinepenalty100%
\else\sloppy\clubpenalty4000\widowpenalty4000\interlinepenalty500\fi%
    \sfcode`\.=1000\relax}

\long\def\@makecaption#1#2{%
\ifx\@captype\@IEEEtablestring%
\par\@IEEEtabletopskipstrut
\else
\@IEEEfigurecaptionsepspace
\fi
\setbox\@tempboxa\hbox{\normalfont\RFICcaptionsize {#1.}\nobreakspace\nobreakspace #2}%
\ifdim \wd\@tempboxa >\hsize%
\setbox\@tempboxa\hbox{\normalfont\RFICcaptionsize {#1.}\nobreakspace\nobreakspace}%
\parbox[t]{\hsize}{\normalfont\RFICcaptionsize\noindent\unhbox\@tempboxa#2}%
\else
\ifCLASSOPTIONconference \hbox to\hsize{\normalfont\RFICcaptionsize\hfil\box\@tempboxa\hfil}%
\else \hbox to\hsize{\normalfont\RFICcaptionsize\box\@tempboxa\hfil}%
\fi\fi
\ifx\@captype\@IEEEtablestring%
\@IEEEtablecaptionsepspace
\else
\fi}

\newlength\tablecaptiontotableskip
\newlength\figuretocaptionskip
\def\@IEEEfigurecaptionsepspace{\vskip\figuretocaptionskip\relax}%
\def\@IEEEtablecaptionsepspace{\vskip\tablecaptiontotableskip\relax}%

\def\abstract{\normalfont%
\@IEEEabskeysecsize\bfseries\textit{\abstractname}\,\bfseries\textit{---}\,%
\@IEEEgobbleleadPARNLSP}%

\def\IEEEkeywords{\normalfont%
\@IEEEabskeysecsize\bfseries\textit{\IEEEkeywordsname}\,\bfseries\textit{---}\,%
\@IEEEgobbleleadPARNLSP}%
\def\endIEEEkeywords{\relax\vspace{0.67ex}%
\par\if@twocolumn\else\endquotation\fi%
\normalsize\normalfont}%

\DeclareRobustCommand*{\RFICauthorrefmark}[1]{\raisebox{0pt}[0pt][0pt]{\textsuperscript{\footnotesize{#1}}}}%
\def\@IEEEauthorblockNtopspace{0ex}
\def\@IEEEauthorblockAtopspace{1mm}
\def\tablename{Table}
\def\thetable{\arabic{table}}
\def\IEEEkeywordsname{Keywords}
\def\subsubsection{\@startsection{subsubsection}{3}{\z@}{1.5ex plus 1.5ex minus 0.5ex}%
{0.7ex plus .5ex minus 0ex}{\normalfont\normalsize\itshape}}%
\def\@seccntformat#1{\csname the#1dis\endcsname\relax}
\def\thesectiondis{\thesection.\hskip 0.5em}
\def\thesubsectiondis{{\hbox to\parindent{\Alph{subsection}.}}}
\def\thesubsubsectiondis{{\hbox to \parindent{\arabic{subsubsection})}}}
\def\theparagraphdis{{\hbox to \parindent{\alph{paragraph})}}}
\IEEEilabelindentA \parindent
\IEEEilabelindent \IEEEilabelindentA
\IEEEelabelindent \parindent
\IEEEdlabelindent \parindent
\IEEElabelindent \parindent

\newlength\@RFICparindent
\newcommand\RFICdisplayacksection[1]{%
\ifIsBlindReviewVersion%
\noindent\phantom{\parbox[t]{\columnwidth}{\normalbaselines\setlength{\parindent}{\@RFICparindent}{#1}\strut}}
\else%
\noindent\parbox[t]{\columnwidth}{\normalbaselines\setlength{\parindent}{\@RFICparindent}{#1}\strut}%
\fi%
}%

\makeatother

\begin{document}
\raggedbottom
%
%
%
\title{A Study on THz Plasmonics in a CMOS Continuum Transistor Array}
%
%
%

\RFICthispaperforblindreview
\RFICthispaperforfinalpublication
\RFICauthor{%
\RFICauthorblockNAME{
H. Naghavi\RFICauthorrefmark{\#1},
S. Chakraborty\RFICauthorrefmark{\#2},
Z. Khalifa\RFICauthorrefmark{\$3},
J. Gruber\RFICauthorrefmark{\$4},
A. V. Muppala\RFICauthorrefmark{\$5},
C. Shen\RFICauthorrefmark{\$6},
D. Sabu \RFICauthorrefmark{\#7},
A. Cathelin\RFICauthorrefmark{$\dagger$8},
E. Afshari\RFICauthorrefmark{\$9}
}
\\%
\RFICauthorblockAFFIL{
\RFICauthorrefmark{\#}University of Washington, Seattle, USA\\
\RFICauthorrefmark{\$}University of Michigan, Ann Arbor, USA\\
\RFICauthorrefmark{$\dagger$}STMicroelectronics, Crolles, France
}
\\%
\RFICauthorblockEMAIL{
\{\RFICauthorrefmark{1}naghavi, \RFICauthorrefmark{2}shrutic2 \RFICauthorrefmark{7}denils\}@uw.edu, \{\RFICauthorrefmark{3}zainkh, \RFICauthorrefmark{4}jgrub, \RFICauthorrefmark{5}mavarma, \RFICauthorrefmark{6}shenchn, \RFICauthorrefmark{9}afshari\}@umich.edu, \RFICauthorrefmark{8}andreia.cathelin@st.com
}
}
%

\maketitle
%
%
%
\begin{abstract}
\label{0-abstract}
This work addresses the limitations of CMOS at terahertz (THz) frequencies, where charge transit time and parasitic capacitances restrict the maximum operating frequency, $f_{\text{max}}$. As transistor dimensions shrink, reduced current handling capabilities further challenge CMOS, necessitating novel circuit design approaches for the THz domain. By leveraging the plasma characteristics of electron channels in CMOS transistors, this study explores a potential solution for THz signal amplification. Key mechanisms in plasma wave amplification within a continuum transistor array (CTA) formed by 28 nm fully depleted silicon-on-insulator (FD-SOI) CMOS transistors are investigated. A hydrodynamic transport model combined with Pierce's theory is presented to describe plasma wave propagation along the CTA. Simulations of gated amplifiers demonstrate the potential for THz signal amplification in advanced fabrication nodes. Finally, a proof-of-concept plasma wave amplifier operating at 700 GHz has been designed and fabricated, exhibiting amplification along the plasma wave propagation path.
\end{abstract}
%
%

\section{Introduction}
\label{1-introduction}
The frequency range from 0.1 to 10 THz, referred to as the THz gap, presents enhanced bandwidth capacity and is, therefore, highly appealing for applications like high-resolution radar imaging, spectroscopy, and high-data-rate communications. Nonetheless, achieving these high-performance systems necessitates effective amplifier modules to mitigate significant signal losses. Conventional CMOS and HBT transistors fail to operate as active devices at THz frequencies due to charge transit time and parasitic capacitances, which limit their $f_{\text{max}}$. Smaller transistor dimensions further reduce breakdown voltage and current handling. With $f_{\text{max}}$ showing minimal improvement over the last years, new circuit design approaches are needed to surpass this limit. Studies suggest that leveraging the plasma nature of electron channels in CMOS transistors could enable fundamental signal generation at the THz band \cite{Otsuji2014}. 

\begin{figure}[t!]
    \centering
    \includegraphics[width=0.95\linewidth]{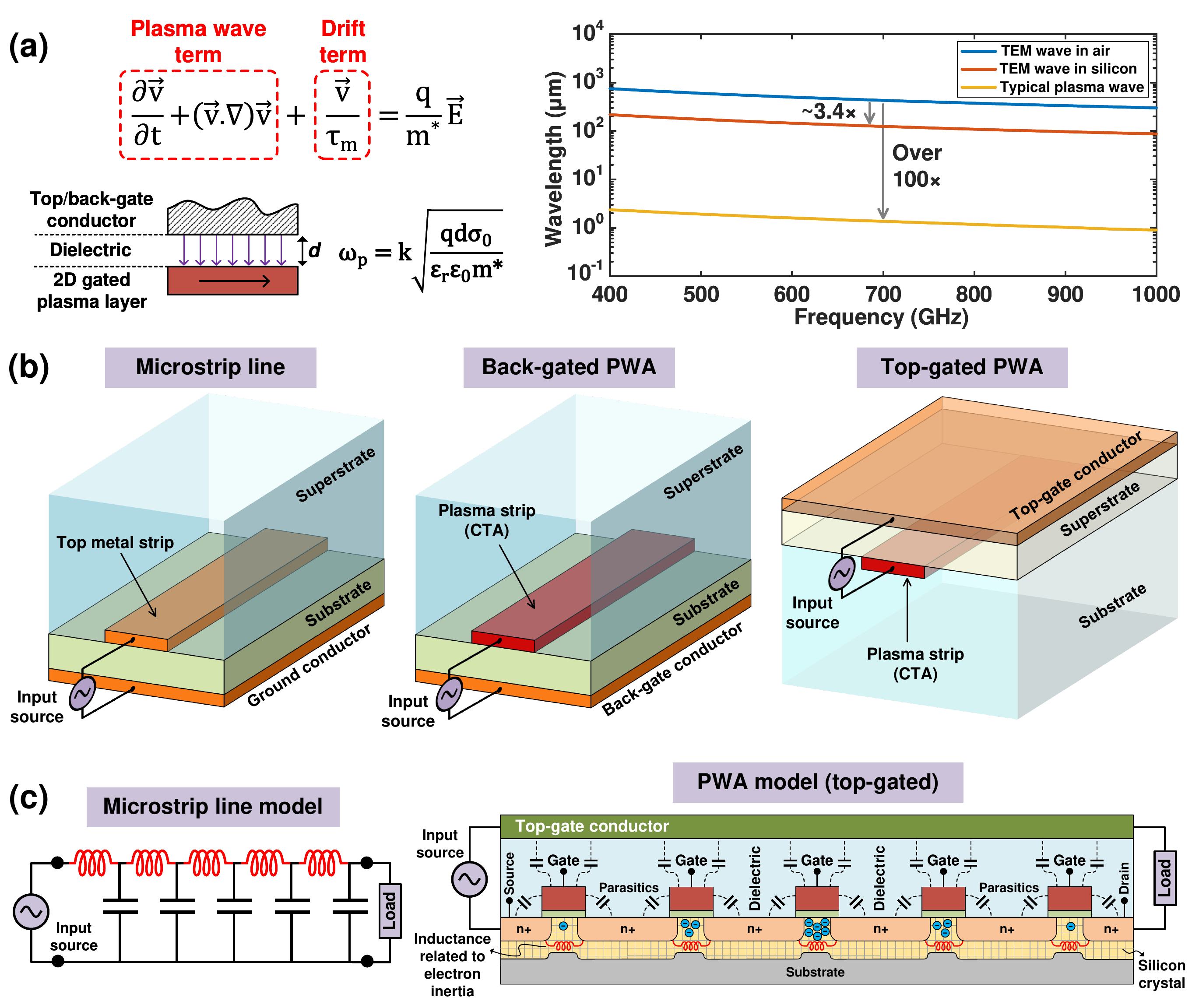}
    \caption{(a) Plasma equation and dispersion diagram. (b) Different forms of plasma wave amplifiers (PWA). (c) PWA modeling.}
    \label{fig:overview}
\end{figure}

Distributed amplifiers \cite{Tarar2024} and traveling-wave transistors \cite{Podgorski1982} incorporate parasitic elements into circuit structures but remain constrained by the charge transit time (drift velocity) in the device channel, approximately $10^7 \, \text{cm/s}$. In contrast, plasma oscillations in field-effect transistors (FETs) can generate plasma waves with velocities on the order of $10^8 \, \text{cm/s}$, significantly exceeding electron drift velocity \cite{Otsuji2014}. Figure \ref{fig:overview}(a) presents the simplified governing equations for electron drift motion, plasma waves, and the plasma dispersion relationship in the device channel. Key parameters include electron velocity ($\Vec{v}$), momentum relaxation time ($\tau_m$), electric field ($\Vec{E}$), electron charge ($q$), effective electron mass ($m^*$), dielectric permittivity ($\epsilon_r \epsilon_0$), wavenumber ($k$), and surface charge density ($\sigma_0$). At THz frequencies, the nonlinear term 
($(\vec{v} \cdot \nabla) \vec{v}$)  transfers energy from the DC electron stream to plasma waves, rendering the drift term negligible. Additionally, Fig. \ref{fig:overview}(a) compares plasma wavelengths with electromagnetic (E/M) wavelengths, showing that plasma wavelengths in semiconductors are over 100 times smaller, making plasma waves highly suitable for THz integrated electronics.

A body of research on plasma waves in compound semiconductors has been conducted, focusing on the design of mm-wave solid-state traveling-wave amplifiers (SSTWA) \cite{Anastasiadis2023, Pousi2008, Lyubchenko1994}; however, the design of THz plasma wave amplifiers (PWA) that are compatible with standard CMOS processes remains to be explored. In this article, a novel plasma medium, formed by an array of 28 nm FD-SOI CMOS transistors and referred to as the ``continuum transistor array (CTA),'' is introduced. The theory, simulation, and measurement results of two types of novel THz PWA (back- and top-gated) utilizing CTA as their active medium are reviewed, as shown in Fig. \ref{fig:overview}(b). The distinction between these amplifiers and a conventional microstrip line is elucidated in Fig. \ref{fig:overview}(c), where the strip line for the THz PWA is constructed from CTA. Additionally, the inductances formed in the PWA arise from the electron inertia in the transistor channels. Overall, this work provides a new understanding of plasmonic effects in CMOS devices operating in the THz band.

\section{Theory}
\label{2-theory}

\begin{figure}
    \centering
    \includegraphics[width=0.9\linewidth]{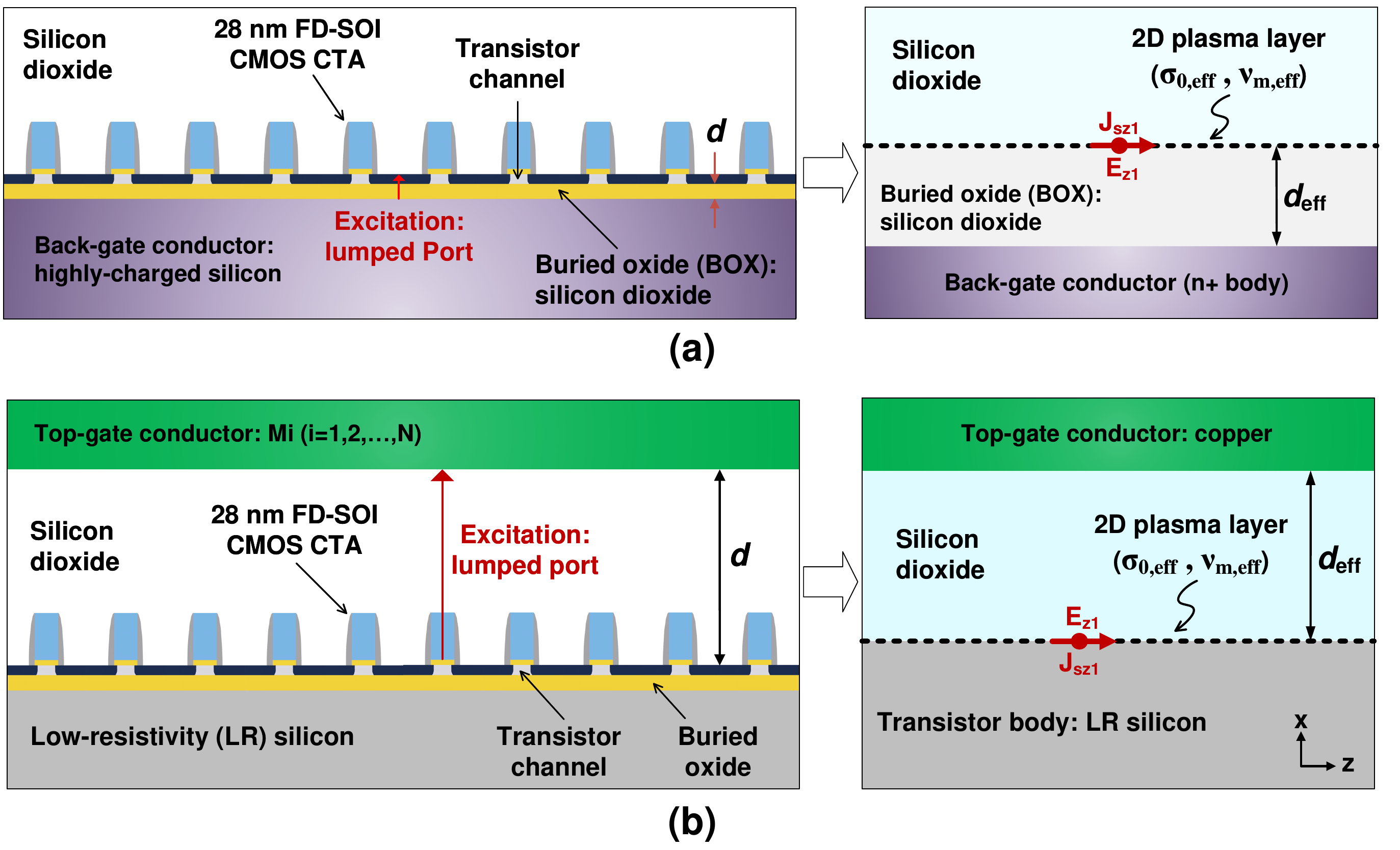}
    \caption{Equivalent 2D plasma layer for (a) back-gated and (b) top-gated CTA.}
    \label{fig:equivalent_plasma}
\end{figure}

To understand the CTA operation mechanism, it is essential to analyze gated plasma wave behavior, as illustrated in Fig. \ref{fig:equivalent_plasma} for back-gated and top-gated topologies. To avoid confusion, it should be emphasized that the top- and back-gate conductors are different from the transistor gate contacts. In this simple model, the CTA is represented by a 2D plasma layer with an effective surface charge density $\sigma_{0, \text{eff}}$ and a charge collision frequency $\nu_{m, \text{eff}}$. The variables $d$ and $d_{\text{eff}}$ are the distances between the back- and top-gate conductors with the CTA and the equivalent plasma layer, respectively. The key considerations here are to understand the relationship between a single transistor's parameters and the equivalent 2D plasma layer parameters. Plasma waves, analogous to space-charge waves in microwave tubes, require the solution of the equations of motion, continuity, and Maxwell’s equations to fully describe their behavior. This analysis reveals how energy from the DC electron stream is transferred to plasma waves, which are further explained below.

\label{2.1-plasma-physics-hydrodynamic-model}
\subsection{Plasma Physics: Hydrodynamic Model}

The hydrodynamic transport model is used to analyze electron motion in transistor channels due to the computational intensity required to solve the Boltzmann Transport Equation. This model simplifies the analysis by focusing on the current-continuity and momentum conservation equations. Equation (\ref{hydrodynamic-transport-eqns-2}) represents the momentum conservation in the hydrodynamic transport model, which is the core of this analysis. Here, $\vec{J}$ is the current density, $\rho$ is the volume charge density, $\nu_m = 1/\tau_m$ is the electron collision frequency, and $v_{\text{therm}}$ is the thermal velocity of charges in the semiconductor \((v_{\text{therm}} = \sqrt{3k_B T/m^*})\).

\begin{equation}
    \frac{\partial \Vec{J}}{\partial t} + (\Vec{v} \cdot \nabla)\Vec{J} + \Vec{J} \nabla \cdot \Vec{v} + \nu_{m} \Vec{J} + \frac{v^2_{\text{therm}}}{2}\nabla \rho = \frac{q\rho}{m^{*}}\Vec{E}
    \label{hydrodynamic-transport-eqns-2}
\end{equation}

For small-signal AC analysis, equations \ref{hydrodynamic-transport-eqns-2} and the current-continuity equation are combined and linearized, enabling the study of plasma waves. Equation (\ref{velocity-sat-regime-transcendental-eqn}) represents the time-harmonic model of the 2D plasma medium at angular frequency $\omega$ ($\frac{\partial}{\partial t}\rightarrow -j\omega$), assuming that the electrons in the plasma layer move with a DC velocity $v_{\text{beam}}$ in the $z$ direction and experience a DC electric field $E_{z0}$. Here, $k_z$ is the $z$-directed wavenumber ($\nabla \rightarrow jk_z \hat{z}$). The DC electric field can be related to the drain-source voltage $V_{DS}$ of any transistor as $E_{z0} \approx \frac{V_{\text{DS}}}{L_g}$, where $L_g$ is the device gate length. The unknowns in this relation are the AC surface current density $J_{sz1}$ and the AC electric field $E_{z1}$. This equation numerically demonstrates that plasma waves facilitate the transfer of energy from the DC terms ($v_{\text{beam}}$ and $E_{z0}$) to the AC terms ($J_{sz1}$ and $E_{z1}$), thereby enabling THz wave amplification. Furthermore, having a lower collision frequency $\nu_{m,\text{eff}}$ is an essential requirement for plasma wave amplification.

\begin{multline}
    \Big[ (v_{\text{beam}}^{2} - \frac{1}{2} v_{\text{therm}}^{2}) k_{z}^{2} - \left(2 \omega v_{\text{beam}} + j \frac{q}{m^{*}} E_{z0}\right) k_{z} \\
    + \left(\omega^{2} + j \nu_{m, \text{eff}} \omega\right) \Big] J_{sz1} = j \omega \frac{q \sigma_{0, \text{eff}}}{m^{*}} E_{z1}
    \label{velocity-sat-regime-transcendental-eqn}
\end{multline}

\begin{figure}[t!]
    \centering
    \includegraphics[width=1\linewidth]{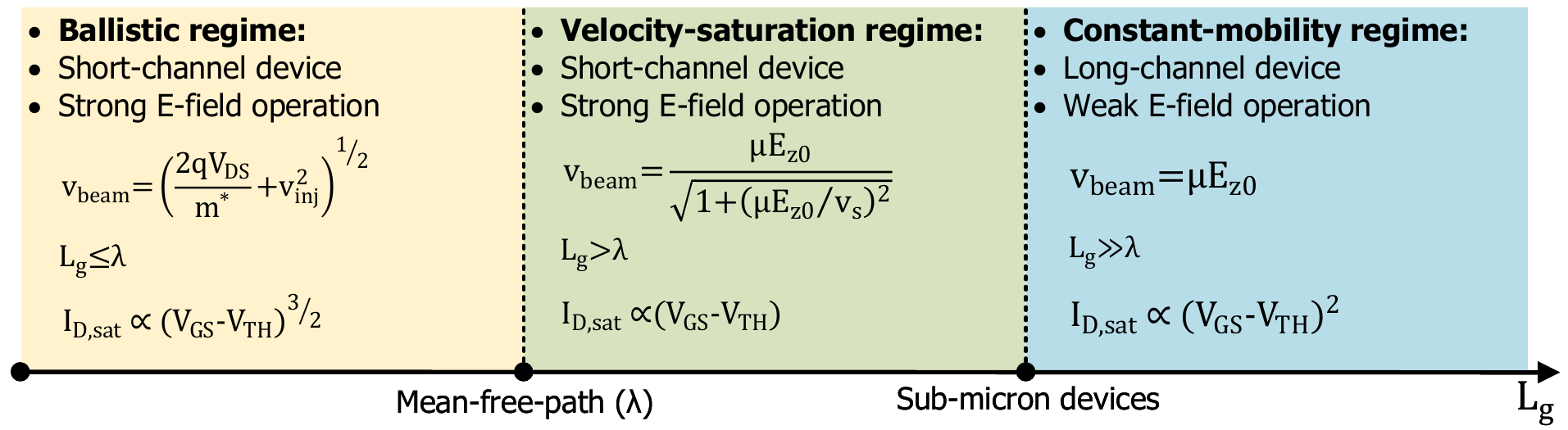}
    \caption{Transistor Operation Regimes}
    \label{fig:transistor-operation-regimes}
\end{figure}

One more step is to relate the DC terms of $v_{\text{beam}}$ and $E_{z0}$. As shown in Fig. \ref{fig:transistor-operation-regimes}, transistor operations are categorized into three main regimes depending on the device gate length ($L_g$): constant mobility, velocity saturation, and ballistic transport. The likelihood of plasma wave amplification is higher in the ballistic regime due to a stronger DC electric field in the device channel and reduced carrier scattering, where the channel length $L_g$ is less than the mean free path $\lambda$ of electrons in the device channel. Research on ultra-small MOSFETs (<100 nm) suggests that reduced carrier scattering in intrinsic silicon channels (as in 28 nm FD-SOI CMOS) may enable ballistic transport, allowing electrons to move with minimal collisions, thereby forming an electronic plasma that supports plasma waves.
\label{2.2-pierce-theory}
\subsection{Pierce Theory}

With the hydrodynamic model of electron transport in a 2D plasma layer established, one more equation is needed to complete the analysis by solving the Maxwell equations shown in Fig. \ref{fig:equivalent_plasma}. Equation (\ref{eq:slow-wave-assumption-acceptable-solutions}) under slow-wave assumptions for plasma waves provides the extra relationship between $J_{sz1}$ and $E_{z1}$ to complete the set of equations (Pierce approach). Here, $\epsilon_a = \epsilon_{ra} \epsilon_0$, $\epsilon_b = \epsilon_{\text{rb}} \epsilon_0 + j\frac{\sigma_{\text{cb}}}{\omega}$, and $\sigma_{\pm} = \text{sign} \{ \text{Re} \{ k_z \} \} = \pm 1$ for forward and backward propagating waves, respectively, where $k_z$ is the $z$-directed complex propagation constant and only $ \text{Re} \{ \sigma_{\pm} k_z\} > 0 $ are acceptable solutions.

\begin{equation}
    E_{z1} \approx \frac{k_z}{j \omega \big(\epsilon_a \coth{(k_z d)} + \epsilon_b \sigma_{\pm}\big)} J_{sz1}
    \label{eq:slow-wave-assumption-acceptable-solutions}
\end{equation}

Due to the small values of metal height ($d, d_{\text{eff}} < 1 \, \mu m$), $k_z d \ll 1$, and $\coth(k_z d) \approx \frac{1}{k_z d}$, the transcendental equation for the gated plasma waves becomes the following:

\begin{scriptsize}
\begin{multline}
     \left[(v_{\text{beam}}^2 - \frac{1}{2} v_{\text{therm}}) k_z^2 - \left(2 \omega v_{\text{beam}} + j \frac{q}{m^{*}} E_{z0} + \frac{q \sigma_{0, \text{eff}}}{m^{*}} \frac{\sigma_{\pm}}{(\epsilon_{a} + \epsilon_{b})}\right) k_z \right. \\ 
     \left. + \left(\omega^2 + j \nu_{m, \text{eff}} \omega\right) \right] (\epsilon_{a} + \epsilon_{b} d_{\text{eff}} \sigma_{\pm} k_z) - \frac{q \sigma_{0, \text{eff}} d_{\text{eff}}}{m^{*}} k_z^2 = 0
     \label{eq:final-form-transcendental-eqn-ballistic-regime-2DEG-gated-plasma-waves}
\end{multline}
\end{scriptsize}

Equation \ref{eq:final-form-transcendental-eqn-ballistic-regime-2DEG-gated-plasma-waves} represents the transcendental equation, which is a third-order polynomial that needs to be solved twice for different values of $\sigma_{\pm}$. For each value of $\sigma_{\pm}$, three $k_z$ values are found, two of which fail to satisfy the presumed conditions of $\text{Re}\{\sigma_{\pm} k_z\} > 0$ and $k_z d \ll 1$. Hence, the analysis yields two final solutions for $k_z$, which correspond to forward and backward propagating plasma waves. A positive value for $\text{Re}\{jk_z\}$ indicates a growing plasma wave along the CTA.

\section{Simulations}
\label{3-simulations}

\begin{figure}[t!]
    \centering
    \includegraphics[width=0.8\linewidth]{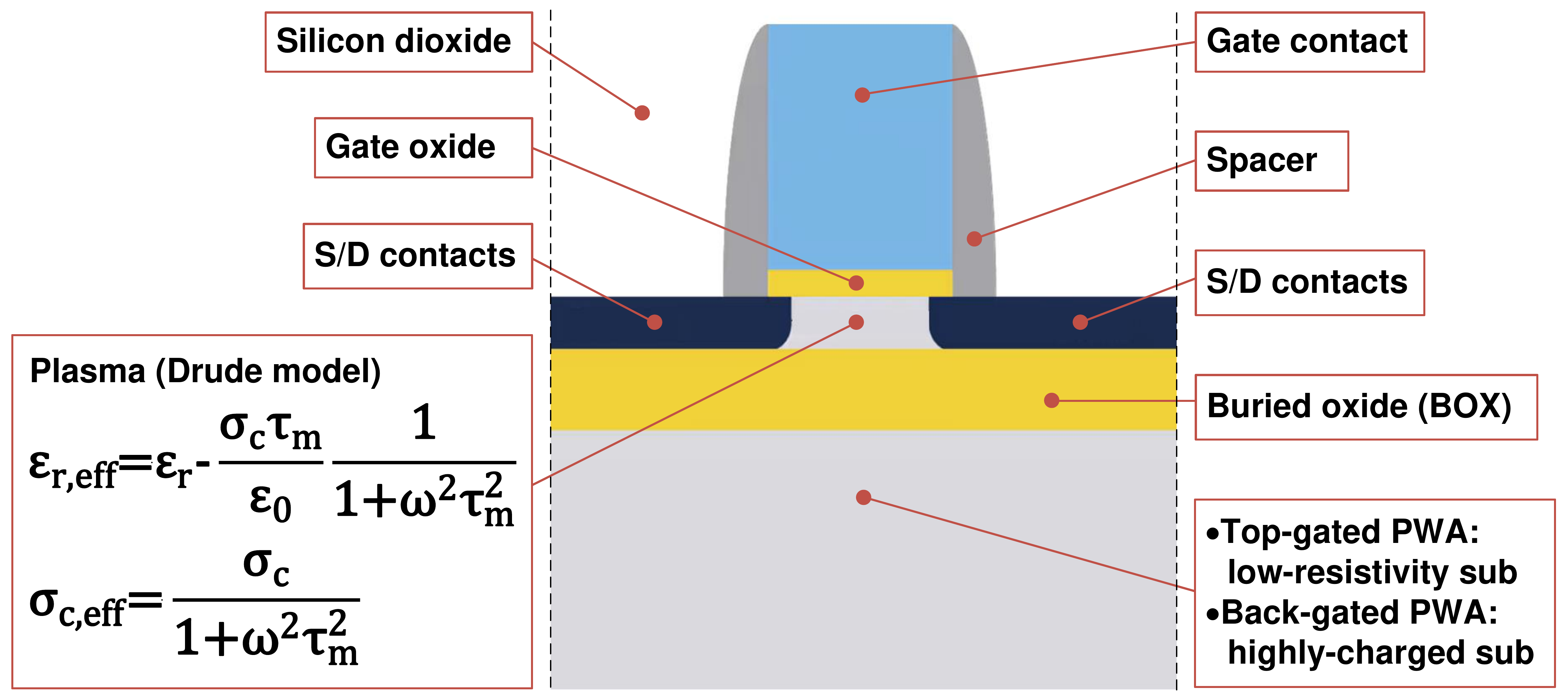}
    \caption{28 nm FD-SOI CMOS transistor incorporating the Drude model. The parameters for each part are not disclosed due to confidentiality.}
    \label{fig:ST-xtor-parameters-drude}
\end{figure}

\begin{figure}[t!]
    \centering
    \includegraphics[width=0.85\linewidth]{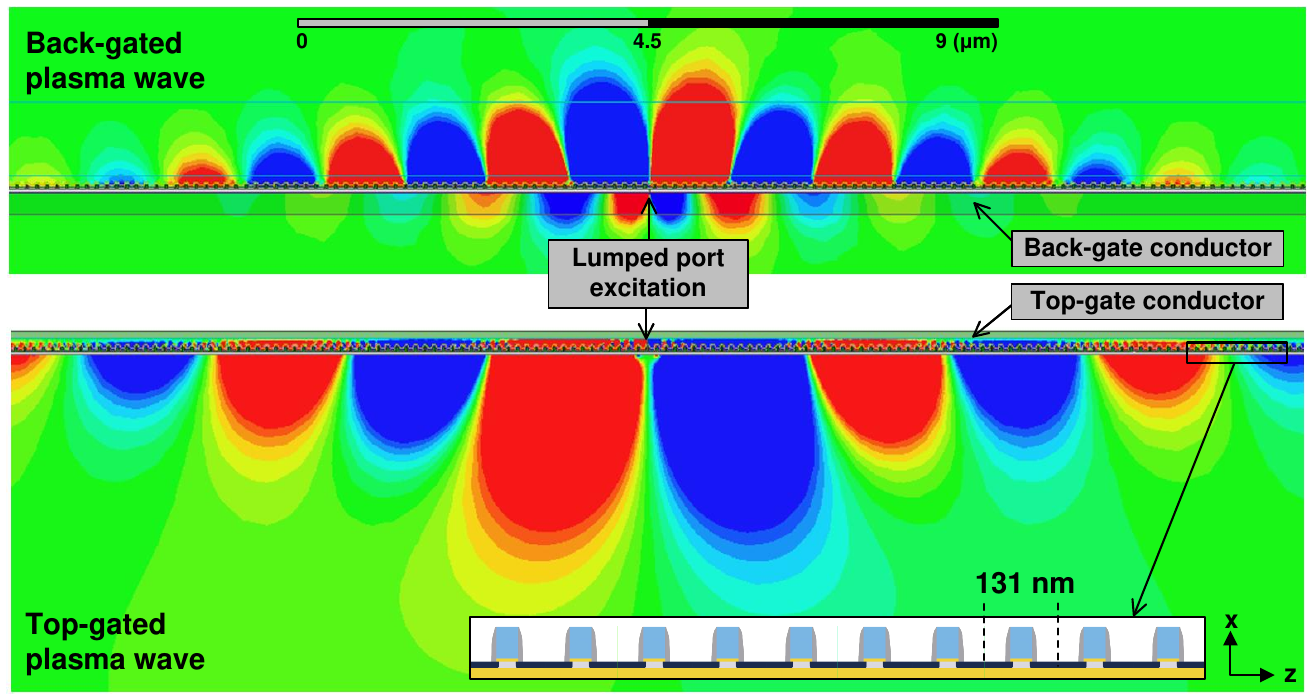}
    \caption{HFSS simulation of plasma wave propagation along the gated CTA at 700 GHz for: $\sigma_c=1.12\times 10^5~(S/m)$, $\mu=0.14~(m^{2}/V\cdot s)$.}
    \label{fig:HFSS-plasma-propagation-xtor-array-ballistic}
\end{figure}

The transcendental equations for gated plasma waves are now established and can be applied to model a CTA. In the absence of DC current in the CMOS transistor ($ V_{\text{DS}} = 0 $), the Drude model can be utilized instead of the hydrodynamic transport model for the plasma region of each transistor. With this simplification, the transistor array can be simulated in HFSS, as illustrated in Fig. \ref{fig:ST-xtor-parameters-drude}, where $ \sigma_{c} $ is the DC conductivity given by $ \sigma_{c} = \frac{q \rho_{0}}{\nu_{m} m^{*}} $, and $ \rho_{0} $ is the DC volume charge density of the device channel. This simplification is reasonable when the DC drift velocity $ v_{\text{beam}} $ of charges in the transistor channel is an order of magnitude less than the plasma wave velocity. Therefore, the calculated values for $ \sigma_{0, \text{eff}} $ and $ v_{m, \text{eff}} $ under zero DC current can be extended to subsequent analyses, even when $ V_{\text{DS}} \neq 0 $ or $ v_{\text{beam}} \neq 0 $ in the theoretical model. The goal is to replace the transistor array with an equivalent 2D plasma layer, as illustrated in Fig. \ref{fig:equivalent_plasma}. The parameter that needs to be extracted from HFSS modeling is the plasma wavenumber $ k_z $. Using this, and (\ref{eq:sigma-nu}), the remaining parameters $ \sigma_{0, \text{eff}} $, $ v_{m, \text{eff}} $, and $ d_{\text{eff}} $ can be determined for the equivalent 2D plasma layer, as shown in Fig. \ref{fig:equivalent_plasma}.

\begin{subequations}
    \begin{equation}
        \sigma_{0, \text{eff}} = m^{*} \omega^{2}\bigg/\operatorname{Re}\left\{ \frac{q k_{z}}{\epsilon_{a} \coth(k_{z} d_{\text{eff}}) + \epsilon_{b}} \right\}
        \label{simplified-trancendental-gated-2deg-eqn-sigma}
    \end{equation}
    
    \begin{equation}
        \nu_{m, \text{eff}} = \frac{q \sigma_{0, \text{eff}}}{m^{*} \omega} \operatorname{Im}\left\{ \frac{k_{z}}{\epsilon_{a} \coth(k_{z} d_{\text{eff}}) + \epsilon_{b}} \right\}
        \label{simplified-trancendental-gated-2deg-eqn-nu}
    \end{equation}
    \label{eq:sigma-nu}
\end{subequations}

\begin{figure}[t!]
    \centering
    \includegraphics[width=0.95\linewidth]{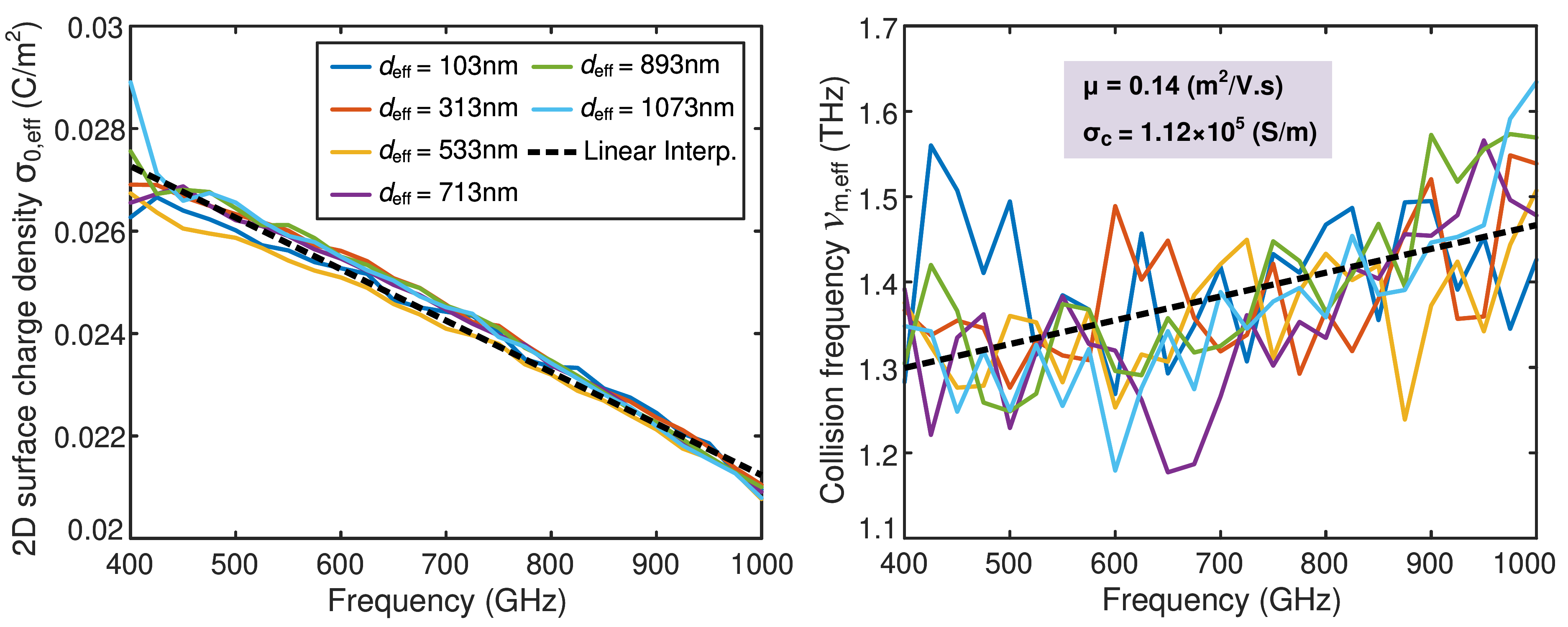}
    \caption{$\sigma_{0, \text{eff}}$ and $\nu_{m, \text{eff}}$ versus frequency for different $d_{\text{eff}}$.}
    \label{fig:effective-n-nu-vs-freq-ballistic}
\end{figure}

\begin{figure}[t!]
    \centering
    \includegraphics[width=0.95\linewidth]{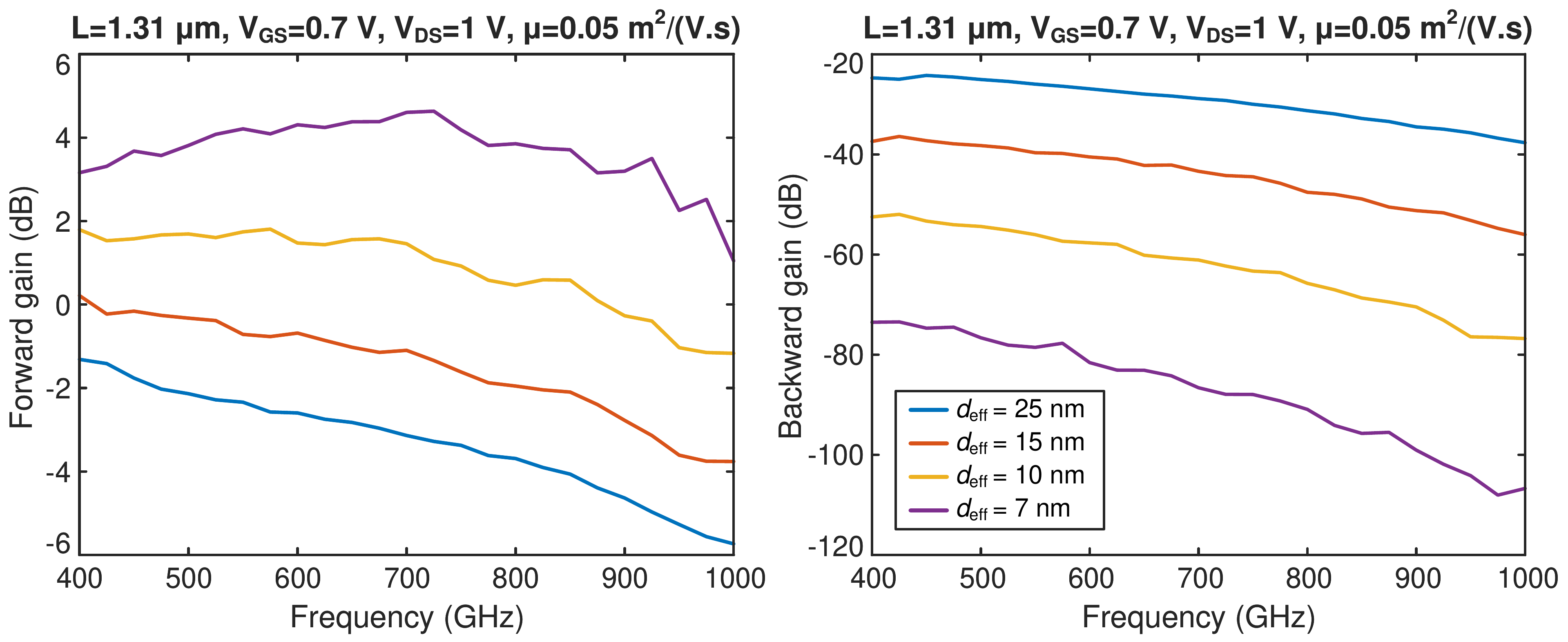}
    \caption{Forward and backward modes of a back-gated PWA.}
    \label{fig:forward-backward-gain}
\end{figure}

\begin{figure}[t!]
    \centering
    \includegraphics[width=0.95\linewidth]{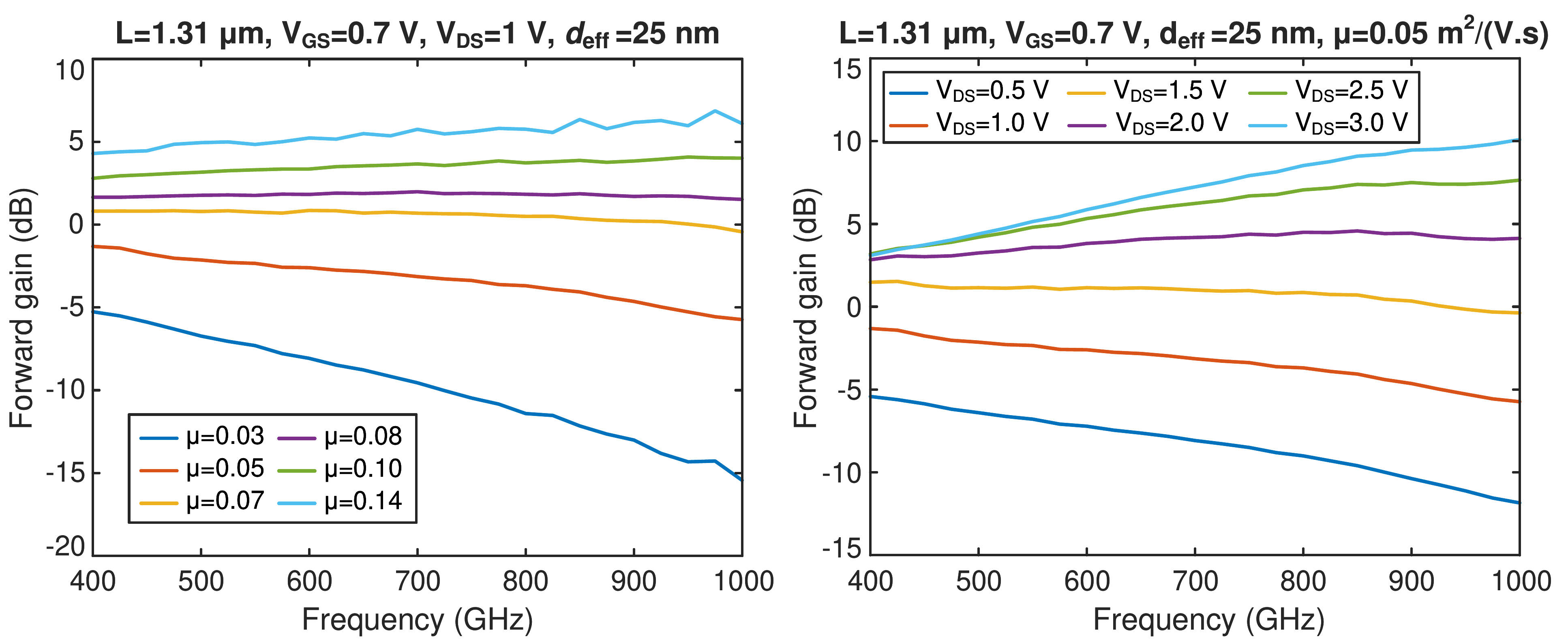}
    \caption{Forward mode of a back-gated PWA.}
    \label{fig:forward-gain-vs-mu-vds}
\end{figure}

\begin{figure}[t!]
    \centering
    \includegraphics[width=0.83\linewidth]{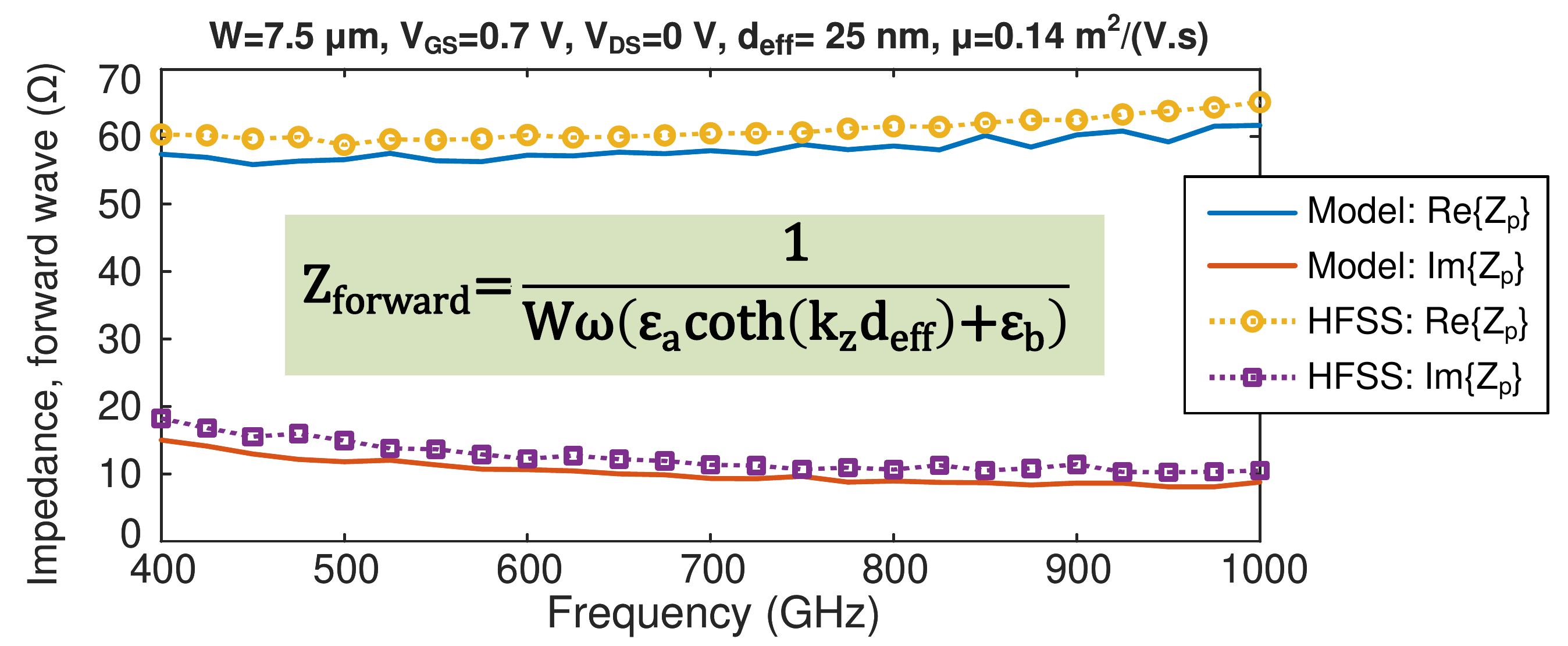}
    \caption{Forward mode impedance of a back-gated PWA.}
    \label{fig:comparison-re-im-impedance-HFSS-vs-developed-model}
\end{figure}

\vspace{0 pt}
HFSS simulations of plasma wave propagation ($E_z$) in back- and top-gated CTA are illustrated in Fig. \ref{fig:HFSS-plasma-propagation-xtor-array-ballistic}. In the HFSS Drude model for the device channel, electron mobility $ \mu = \frac{q\tau_m}{m^*} $ is used in place of $ \tau_m $. For better representation, bulk silicon mobility, $ \mu = 0.14 \, \text{m}^2/\text{V}\cdot \text{s} $ is utilized. Additionally, $ \sigma_c = 1.12\times 10^5 \, \text{S/m} $ corresponds to a gate-source bias of $ V_{\text{GS}} = 0.7 \, \text{V} $. Fig. \ref{fig:effective-n-nu-vs-freq-ballistic} shows the extracted $ \sigma_{0, \text{eff}} $, $ v_{m, \text{eff}} $, and $ d_{\text{eff}} $, with a linear interpolation represented by a dotted line. Using these effective parameters, Eq. (\ref{eq:final-form-transcendental-eqn-ballistic-regime-2DEG-gated-plasma-waves}) can be solved to determine the forward and backward modes of a gated PWA. Figure \ref{fig:forward-backward-gain} depicts the gain for both modes in a back-gated PWA at various $ d_{\text{eff}} $ values assuming $ \mu = 0.05 \, \text{m}^2/\text{V}\cdot \text{s} $ which is approximately the actual electron mobility in the 28 nm FD-SOI CMOS channel. Here, $L= 1.31 \mu m$ is the CTA length made of ten transistors. The results indicate that backward waves cannot deliver gain under any condition, whereas forward waves with $ d_{\text{eff}} < 10 \, \text{nm} $ achieve gain at THz frequencies. Further analysis of a back-gated PWA under varying mobility  $ \mu $ and $ V_{\text{DS}} $ conditions for $ d_{\text{eff}} $ = 25 nm is presented in Fig. \ref{fig:forward-gain-vs-mu-vds}, demonstrating that the PWA can provide fundamental gain at THz frequencies. Additionally, the theoretical and simulated plasma wave impedance for the forward mode in a back-gated PWA are shown in Fig. \ref{fig:comparison-re-im-impedance-HFSS-vs-developed-model}. This impedance is adjustable using the device width $W$.

\section{Experiments}
\label{4-experiments}

\begin{figure}
    \centering
    \includegraphics[width=0.72\linewidth]{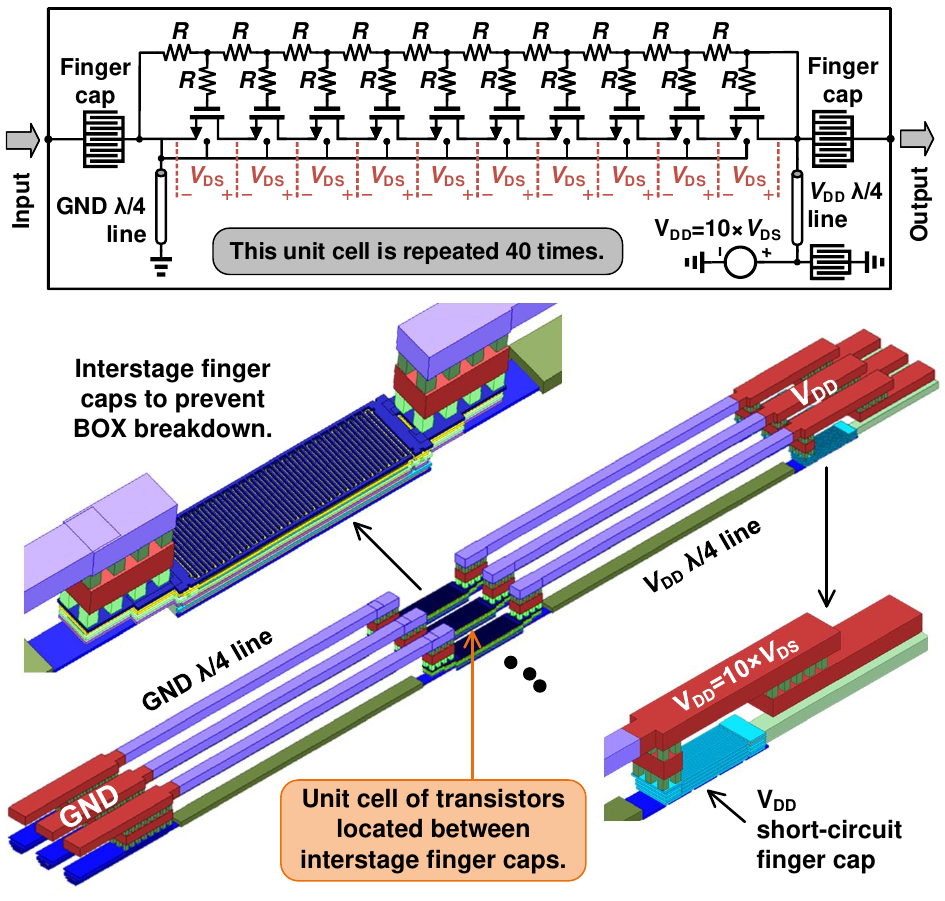}
    \caption{Schematic of PWA made of ten 28 nm FD-SOI CMOS transistors. Interstage finger caps and $\lambda/4$ $V_{\text{DD}}$ and GND lines for back-gated PWA.}
    \label{fig:interstage-finger-caps-lambda-over-VDD-GND-lines-back-gated}
\end{figure}

\begin{figure}[t!]
    \centering
    \includegraphics[width=0.87\linewidth]{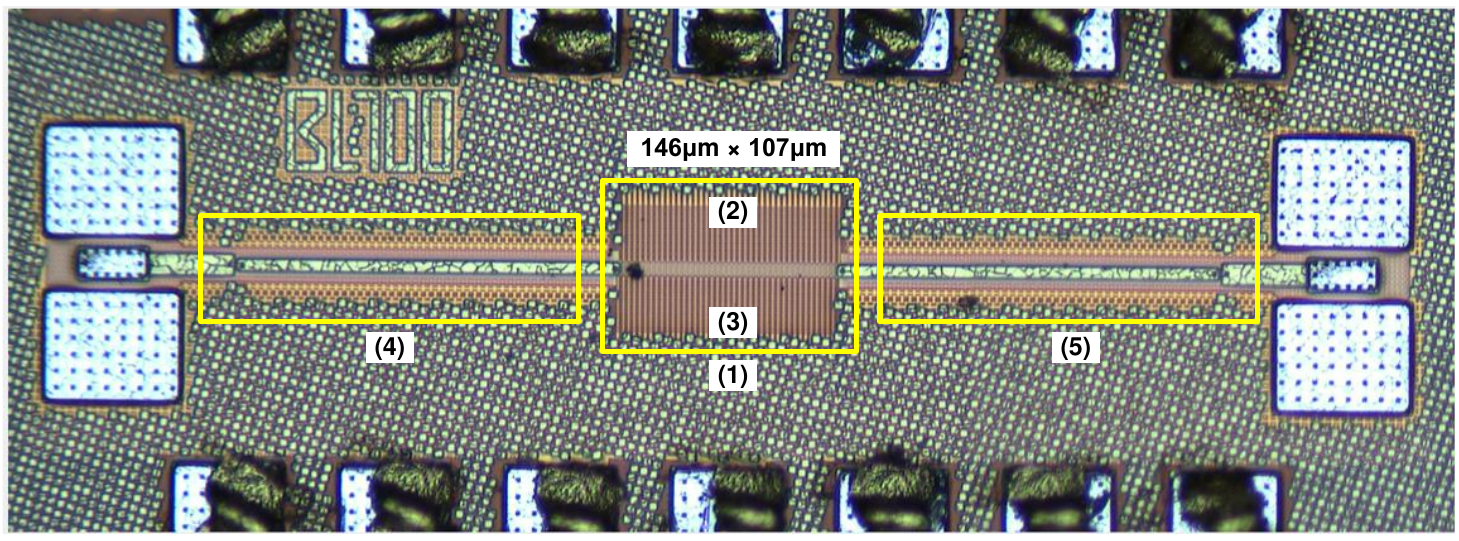}
    \caption{Die micrograph: 1) back-gated PWA; 2,3) $V_{\text{DD}}$ and GND quarter-wave bias lines; 4,5) input and output matching networks.}
    \label{fig:chip-BL700}
\end{figure}

\begin{figure}[t!]
    \centering
    \includegraphics[width=1\linewidth]{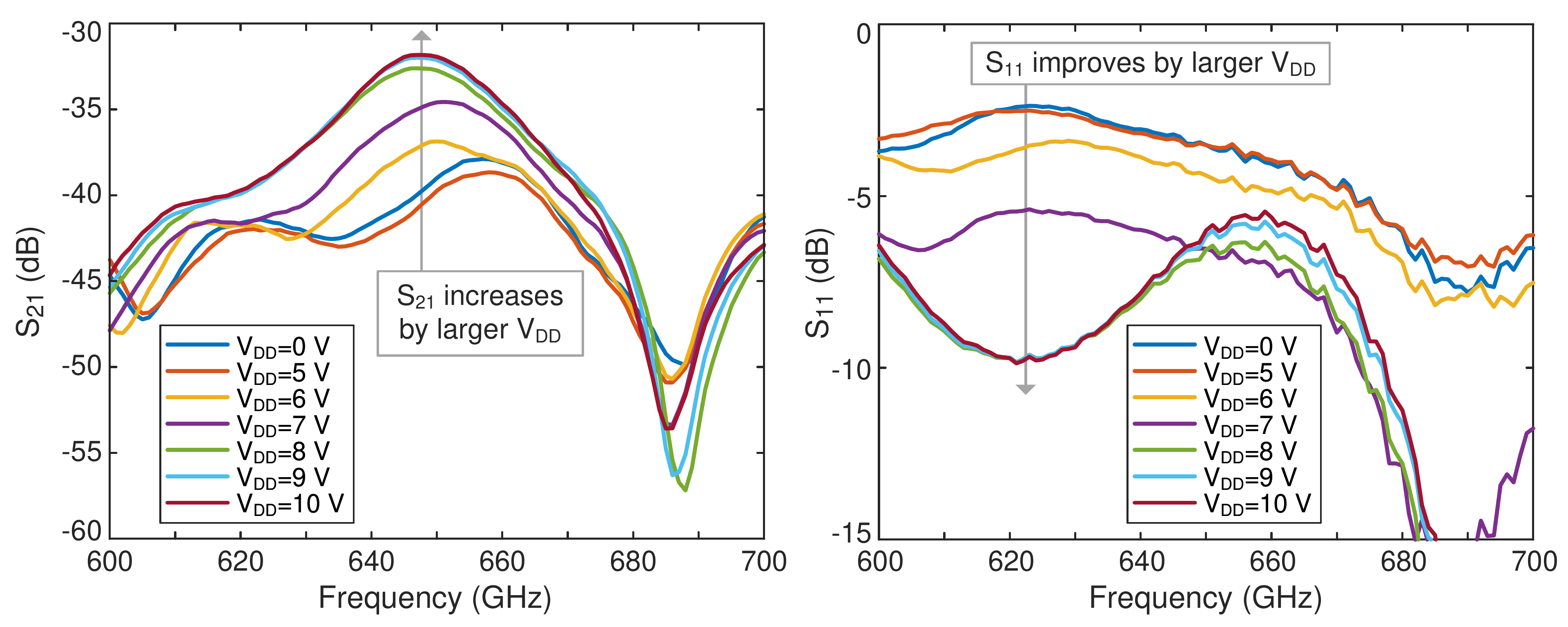}
    \caption{Measured S-parameter results for a back-gated PWA.}
    \label{fig:measured-s21-BL700-v2-amplifier}
\end{figure}

Figure \ref{fig:interstage-finger-caps-lambda-over-VDD-GND-lines-back-gated} shows the schematic and layout of the fabricated back-gated PWA designed for 700 GHz. Ten transistors are biased to maintain a uniform $ V_{\text{GS}} $ and $ V_{\text{DS}} $. Kilo-ohm resistors $R$ and two quarter-wave lines at $ V_{\text{DD}} $ and GND bias the CTA, while preventing THz energy loss through the biasing points. Large interstage finger caps are used at the input/output ports to separate each CTA group, preventing substrate and BOX breakdown. A maximum of 10 V is allowed to be applied across each CTA. Forty of these unit cells are cascaded to form a PWA that achieve sufficient gain. Figure \ref{fig:chip-BL700} illustrates the micrograph of the fabricated back-gated PWA with a total area of 0.31 $\text{mm}^2$ (pad included) and a measured maximum DC power consumption of 1.29 W. Figure \ref{fig:measured-s21-BL700-v2-amplifier} shows the measured S-parameters of the back-gated PWA designed for 700 GHz using FormFactor T750-GSG-25-BT GSG probes, Keysight N5227B-401 PNA, and VDI WM-380 frequency extenders. The FormFactor 172-885 impedance standard substrate is used for calibration. Based on Fig. \ref{fig:forward-gain-vs-mu-vds}, achieving net gain from the CTA requires $V_{\text{DS}} > 2$V, which is constrained by device breakdown limitations. However, an increase in $V_{\text{DS}}$ is expected to enhance $S_{\text{21}}$, as shown in Fig. \ref{fig:measured-s21-BL700-v2-amplifier}. Realizing net gain remains challenging and may necessitate thinning the BOX to less than 10 nm. The measured $S_{\text{11}}$ results are also presented in Fig. \ref{fig:measured-s21-BL700-v2-amplifier}. The frequency shift from 700 GHz to approximately 650 GHz is attributed to minor inaccuracies in the HFSS material modeling. Finally, Table \ref{table:ComparisonTable} provides a comparison of these results with state-of-the-art designs.

\begin{table}[t!]
\caption{Comparison table} \vspace{-10pt}
\renewcommand{\arraystretch}{1.25}
\begin{center}
\tiny
\begin{tabular}{|M{0.8cm}|M{1.6cm}|M{0.4cm}|M{0.5cm}|M{0.55cm}|M{0.4cm}|M{0.3cm}|M{0.9cm}|}
  \hline
  \rowcolor{gray!50}
    \hline
    Ref.  &  Active layer (AL) & L ($\mu\text{m}$)$^\text{a}$ & distance (nm)$^\text{b}$& SWS & E-field (V/mm) & Gain (dB/mm) & Frequency \\
    \hline
    \cite{Lyubchenko1994}   &  n-GaAs (AL) & 2000 & 1000 & Dielectic & 120 & 0.4 & ~10 GHz \\
    \hline
    \cite{Pousi2008}        &  GaAs/AlGaAs & 2600 & $\sim$450 & Dielectic & 15 & 0.8 & 76.5 GHz \\
    \hline
    \cite{Anastasiadis2023} &  GaN/AlGaN & 8.3 & $\sim$30 & Metallic & 964 & 21.8  & 75-110 GHz\\
    \hline
    \textbf{This work}  & 28nm FDSOI CMOS & 1.3$^\text{c}$ & 25 & None & 7,600$^\text{d}$ & 62$^\text{e}$ & 650 GHz \\
    \hline
\multicolumn{8}{p{0.96\linewidth}}{{(a) Channel length. (b) Distance between active layer and slow-wave structure (SWS) or back-gate conductor. (c) Made of 10 FDSOI transistors. (d) $V_{\text{DS}}$= 1 V across 131 nm. (e) With interstage finger cap loss.}} \\
\end{tabular}
\end{center}
\label{table:ComparisonTable}
\end{table}

\section{Conclusion}
\label{5-conclusion}

A novel THz plasmonic circuit using 28 nm FD-SOI CMOS CTA is presented. The mechanism of plasma wave propagation in the CTA leverages the concept of back-/top-gated PWA topologies. Simulations and measurements were conducted for frequencies around 700 GHz. Simulation analysis indicates that PWA is capable of providing gain at THz frequencies under certain conditions. Additionally, experimental results demonstrate the achievement of gain improvement along the plasma wave propagation path by biasing CTA. This work marks the first demonstration of a THz plasmonic amplifier on 28 nm FD-SOI CMOS transistors.


\bibliographystyle{IEEEtran}

\bibliography{IEEEabrv,References}

\end{document}